\documentclass[journal,comsoc]{IEEEtran}

\usepackage[T1]{fontenc}% optional T1 font encoding

\usepackage{cite}
\ifCLASSINFOpdf
  \usepackage[pdftex]{graphicx}
  \graphicspath{{./pdf/}{./jpeg/}}
  \DeclareGraphicsExtensions{.pdf,.jpeg,.jpg,.png}
\else
  \usepackage[dvips]{graphicx}
  \graphicspath{{./eps/}}
  \DeclareGraphicsExtensions{.eps}
\fi
\usepackage{amsmath}
\usepackage[cmintegrals]{newtxmath}
\usepackage{array}
\ifCLASSOPTIONcompsoc
  \usepackage[caption=false,font=normalsize,labelfont=sf,textfont=sf]{subfig}
\else
  \usepackage[caption=false,font=footnotesize]{subfig}
\fi
\usepackage{epstopdf}
\usepackage{todonotes}
\usepackage{textcomp, gensymb}
\usepackage{multirow}
\usepackage{bm}
\usepackage{tabu}
\usepackage{float}
\usepackage{booktabs}
\usepackage{dblfloatfix}
\usepackage{cleveref}
\Crefname{figure}{Fig.}{Figs.} % Fix figure refs for IEEE
\usepackage{balance}
\usepackage{bm}
\usepackage{soul}

\usepackage{fancyhdr}
\newcommand{\copyrighttextsize}{\scriptsize}
\newcommand{\squeezeupbeforetitle}{\vspace{-5mm}}
\newcommand{\squeezeupaftertitle}{\vspace{-8.17mm}}
\newcommand{\squeezeupafterindexterms}{\vspace{-2.5mm}}

\newcommand{\squeezeupafterfigure}{\vspace{-1mm}}
\newcommand{\squeezeupbeforetable}{\vspace{-1.5mm}}
\newcommand{\squeezeupaftertable}{\vspace{-1.5mm}}
\newcommand{\squeezeupbetweensubfigs}{\vspace{-1.5mm}}

\newcommand{\hlchanges}{}

\begin{document}

% \title{Foliage Analysis for 28-GHz \\ Millimeter Wave in a Coniferous Forest}
\title{\squeezeupbeforetitle Propagation Modeling Through Foliage \\in a Coniferous Forest at 28 GHz}
% \title{Foliage Analysis for 28-GHz Millimeter Wave Propagation in a Coniferous Forest}

\author{
	\IEEEauthorblockN{
		Yaguang~Zhang, 
		Christopher~R.~Anderson, \\
		%~\IEEEmembership{Senior Member,~IEEE},	\\	
		Nicolo~Michelusi,
		David~J.~Love,
		%~\IEEEmembership{Fellow,~IEEE},
        Kenneth~R.~Baker, 
        %~\IEEEmembership{Senior Member,~IEEE},	
		and James~V.~Krogmeier
		%, ~\IEEEmembership{Member,~IEEE}
	}\squeezeupaftertitle

		\thanks{Y. Zhang, N. Michelusi, D. J. Love, and J. V. Krogmeier are with the School of Electrical and Computer Engineering, Purdue University, West Lafayette, IN 47907, USA (e-mail: \{ygzhang, michelus, djlove, jvk\}@purdue.edu).}
		\thanks{C. R. Anderson is with the Department of Electrical and Computer Engineering, United States Naval Academy, Annapolis, MD 21402, USA (e-mail: canderso@usna.edu).}
		\thanks{K. R. Baker is with the Interdisciplinary Telecommunications Program, University of Colorado Boulder, Boulder, CO 80309, USA (e-mail: ken.baker@colorado.edu).}
		\thanks{Sponsorship for this work was provided by NSF research grant 1642982.}
}

\maketitle

\thispagestyle{fancy}
 
\begin{abstract}
The goal of this article is to investigate the propagation behavior of 28-GHz millimeter wave in coniferous forests and model its basic transmission loss. Field measurements were conducted with a custom-designed sliding correlator sounder. Relevant foliage regions were extracted from high-resolution LiDAR data and satellite images. Our results show that traditional foliage analysis models for lower-frequency wireless communications fail to consistently output correct path loss predictions. Novel fully automated site-specific models are proposed to resolve this issue, yielding 0.9~dB overall improvement and up to 20~dB regional improvement in root mean square errors. 
%A comprehensive model comparison is also provided to elucidate the pros and cons of different modeling approaches.
\end{abstract}


% \fi
\IEEEpeerreviewmaketitle

\squeezeupafterindexterms
\section{Introduction}
\IEEEPARstart{W}{ith} the rapid standardization process of 5G
%communication 
networks~\cite{busari20185g}, millimeter waves (mm-waves) have garnered great attention worldwide from industry, academia, and government. % To incorporate mm-wave bands into future wireless telecommunication networks, one of the biggest challenges 
A major issue is to better understand the propagation characteristics of mm-wave signals.
%and accordingly predict channel states for network planning and operation as needed. 
Many mm-wave channel measurement campaigns have recently been carried out in urban and suburban environments~\cite{rappaport2013millimeter, rappaport2017small, zhang201728ghzicc, zhang2018improving}. 
\hlchanges{
    However, very limited effort has been put into validating and improving current channel models in overcoming vegetation blockage. This is a key element of sensor data collection in forestry and agriculture~\cite{wu2017propagation} for preventing cost incurred by under/over-deployment of the sensors and improving their communication performance.}
In~\cite{papazian1992wideband}, a constant excess path loss of around 25~dB was observed at 28.8~GHz through a pecan orchard for paths with roughly 8 to 20 trees. 
More recent works~\cite{zhang2018improving, rappaport201573} reported low attenuation values per unit foliage depth of 0.07~dB/m at 28~GHz and of 0.4~dB/m with 3~dB deviation at 73~GHz, respectively.
% More recent works \cite{rappaport201573} and \cite{zhang2018improving} reported low attenuation values per unit foliage depth of 0.4~dB/m with a 3~dB deviation at 73~GHz and 0.07~dB/m at 28~GHz, respectively.
%, significantly low compared with currently available models. 
\hlchanges{
    In~\cite{shaik2016millimeter}, attenuation with a dual-slope structure was observed for out-of-leaf measurements at 15~GHz, 28~GHz and 38~GHz in forest environments.}
Moreover, even though a variety of modeling approaches have been considered, most of them ignore site-specific geographic features~\cite{zhang2018improving}.
%, which mm-waves are sensitive to due to blockages.
A comprehensive analysis for attenuation in vegetation is required to validate those observations
%, showcase different modeling approaches, 
and make improvements %appropriate for 
to mm-wave propagation modeling. % if possible, especially because traditional models ignore all or most of site-specific geographic features that mm-waves are sensitive to due to blockages.

\hlchanges{We explore this gap by investigating mm-wave propagation at 28~GHz through a coniferous forest in Boulder, Colorado, where we recorded a total of 1415 basic transmission loss measurements.}
%We explore this research gap by investigating the mm-wave propagation at 28~GHz through vegetation. Using a portable custom-designed sliding correlator sounder, we carried out a measurement campaign in a coniferous forest near Boulder, Colorado and obtained a total of 1415 basic transmission loss measurements. 
A comprehensive model comparison is provided to elucidate the pros and cons of different modeling approaches for predicting signal attenuation through vegetation. Novel site-specific models with consistently better performance than existing models are developed. 

\section{Measurement Setup}

\begin{figure}[t]
	\captionsetup[subfigure]{justification=centering}
	\centering
% 	\subfloat[TX Installed Near the Forest]{
% 		\includegraphics[height=1.20in]{IMG_20180331_120152_TxView_Enhanced_50.jpg}
% 		\label{fig_forest_photo}
% 	}
% 	\hfil
% 	\subfloat[RX Tracks Illustrated with Basic Transmission Loss Results][RX Tracks Illustrated with\\Basic Transmission Loss Results]{
% 		\includegraphics[height=1.21in,trim={0 0.03in 0 0},clip]{./eps/3_1_allContiTracks.eps}
% 		\label{fig_rx_tracks}
% 	}
%     \vspace*{-1mm}
	\subfloat[RX tracks illustrated with basic transmission loss results\squeezeupbetweensubfigs]{
    {\parbox{3in}{\centering\includegraphics[height=2in]{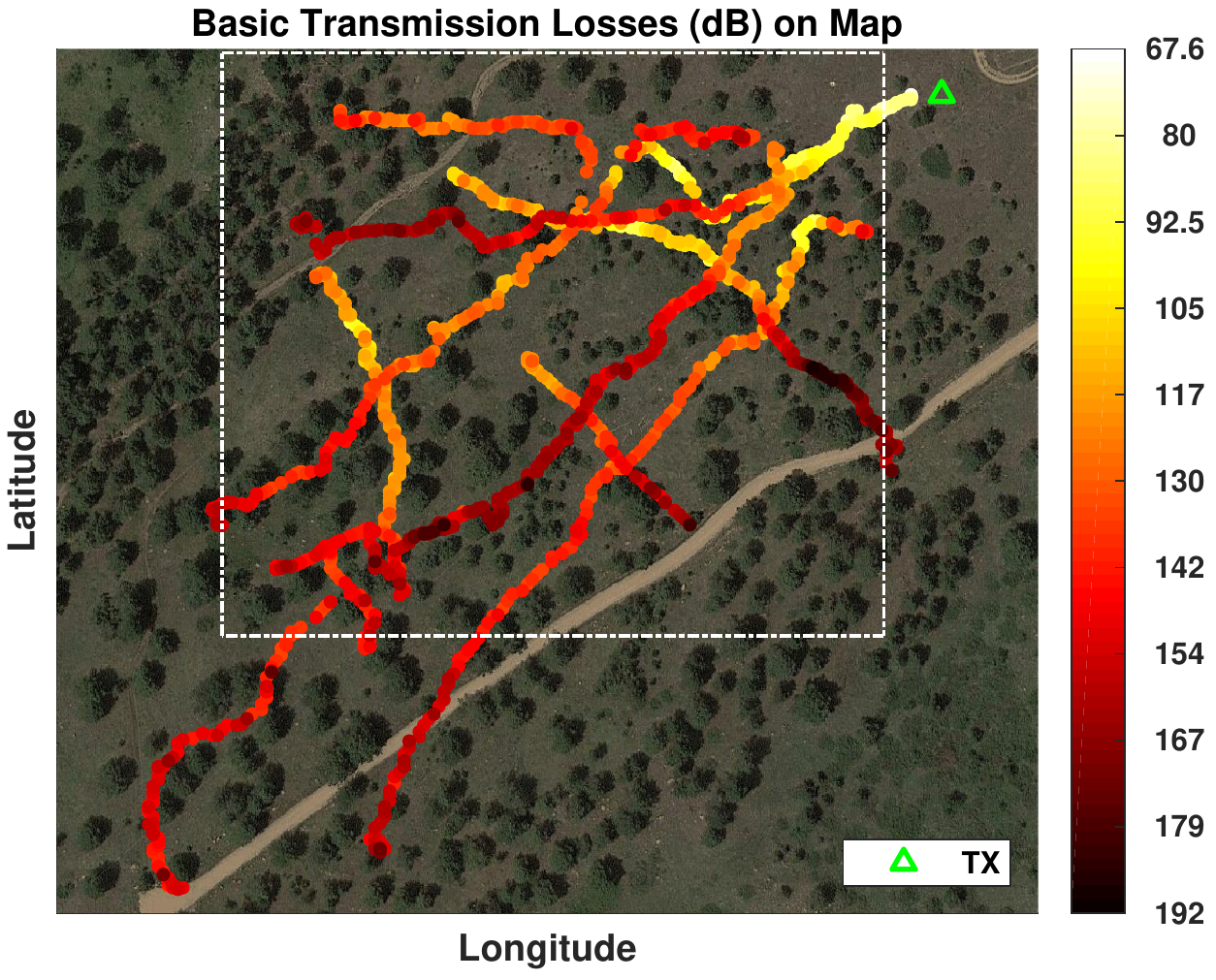}}} % 3_1_allContiTracks_normalSize.eps => zhang_WCL2018-1396_fig1.eps
		\label{fig_rx_tracks}
    }
    \hfil   
	\subfloat[Site-specific information available for the measurement site]{
		{\parbox{3in}{\centering\includegraphics[height=2in,trim={0.01in 0.19in 0 0.19in},clip]{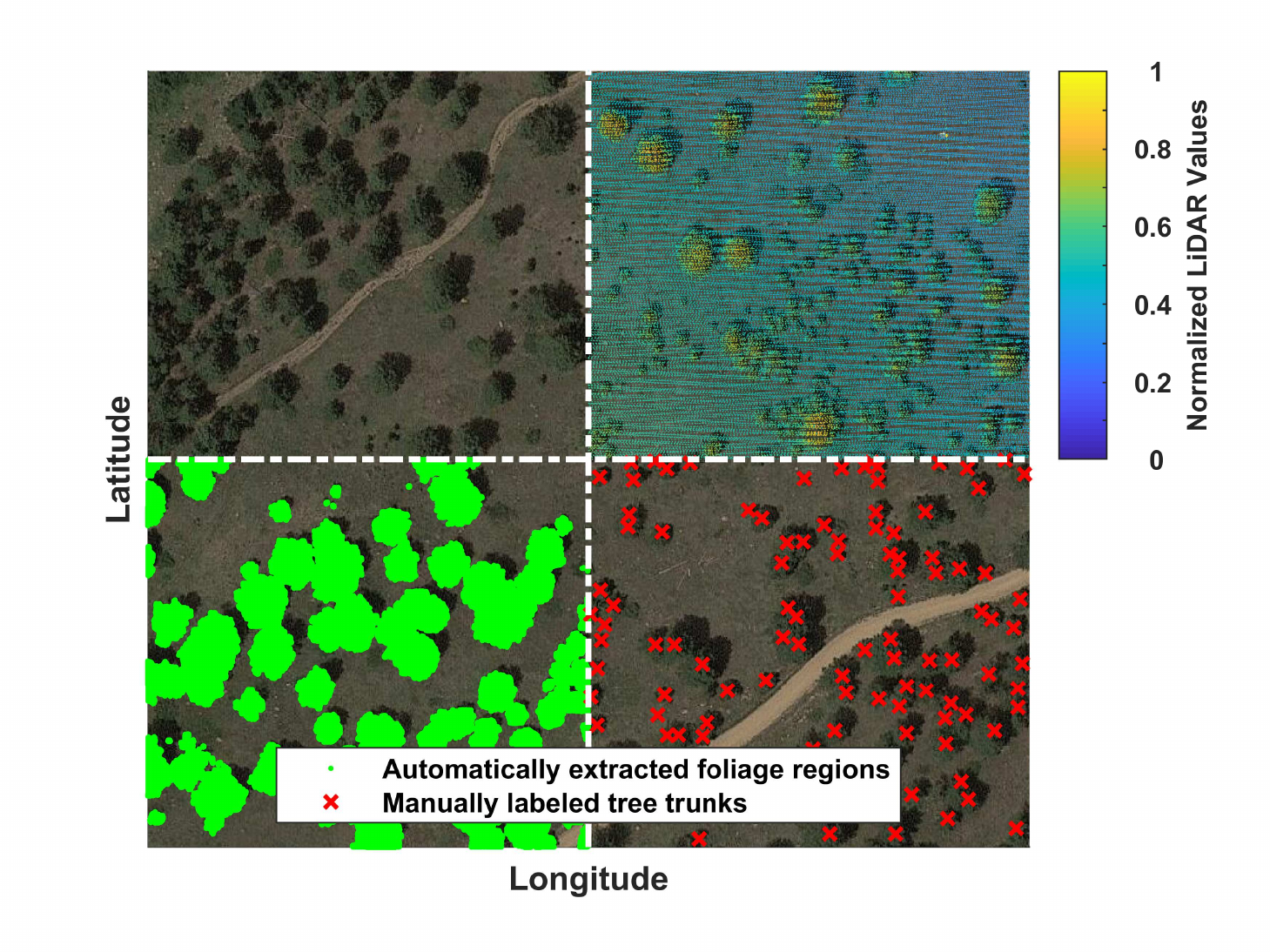}}} % 2_0_Overview_For_Veg_Manual_Matlab_Exp.eps => zhang_WCL2018-1396_fig2.eps
		\label{fig_tree_locaitons_zoomed_in}
	}
	\caption{
	    Overview of the measurement campaign. 
    	(a) The TX was installed at the edge of a coniferous forest. The RX followed 10 different tracks. % in the forest to record the signal continuously. 
        One basic transmission loss result was computed for each second of the recorded signal to match the GPS data. (b) We have zoomed into the dotted-square area in (a) to better illustrate the site-specific features. 
        \hlchanges{
            Overlaid on top of the satellite image are %USGS 
            LiDAR data, %automatically-extracted 
            foliage regions, and %manually-labeled 
            trunk locations, respectively.
            % Satellite images from Google Maps are used here as background. 
    }}
	\label{fig_measurement_overviews}
    \squeezeupafterfigure
\end{figure}

The measurement system in our previous work~\cite{zhang201728ghzicc} was utilized. The receiver (RX), with a chip rate of 399.95 megachips per second, was installed in a backpack and powered by a lithium-ion polymer battery for portability purposes. As illustrated in \Cref{fig_measurement_overviews}\subref{fig_rx_tracks}, the transmitter (TX) was set up at the edge of the forest, while the RX was moved in the coniferous forest to continuously record the signal along with the GPS location information. Basic transmission losses were computed accordingly. The TX antenna was adjusted before each signal recording activity to point to the middle area of the track to be covered. Beam alignment was achieved at the RX side using a compass.
% Relevant foliage regions were extracted from the United States Geological Survey (USGS) LiDAR and terrain elevation data. Tree locations were also manually labeled according to USGS LiDAR data and high-resolution satellite images from Google Maps. These data enabled us to view channel modeling in a site-specific manner. 
% Besides the basic transmission loss measurements and their corresponding GPS data, we also obtained satellite images from Google Maps and LiDAR data from USGS. Tree locations were manually labeled accordingly. And foliage regions were automatically extracted by comparing the LiDAR data with USGS terrain elevation data. 
We also obtained satellite images from Google Maps and LiDAR data from the United States Geological Survey (USGS). Tree locations were manually labeled accordingly. Foliage regions were automatically extracted by comparing the LiDAR data with USGS terrain elevation data. 
These site-specific geographic features of the forest are illustrated in \Cref{fig_measurement_overviews}\subref{fig_tree_locaitons_zoomed_in}. 
\hlchanges{
    Boulder is a semi-arid environment with low humidity and minimal rainfall. Measurements were performed on a warm spring day under mostly sunny conditions.
}

\section{Foliage Analysis for the Coniferous Forest}

We compared three empirical foliage analysis models: the partition-dependent attenuation factor (AF) model~\cite{durgin1998measurements}, the ITU-R obstruction by woodland model~\cite{itu2016vegetation}, and Weissberger's model~\cite{weissberger1982initial}.
%, which take quite different approaches in predicting the attenuation due to vegetation
To tune these models,
%to our dataset and evaluate the potential of our site-specific modeling attempts, 
four parameters were computed for each measurement location: the distance between the TX and the RX, the number of tree trunks within the first Fresnel zone, the foliage depth along the line-of-sight (LoS) path, and the foliage area within the first Fresnel zone. These computations were performed in a three-dimensional (3D) reference system using Universal Transverse Mercator coordinates $(x, y)$ and altitude. Based on these results, site-specific models were introduced to improve path loss predictions.

%% \footnote{Post processing code of our work is available at:\\ https://github.com/YaguangZhang/NistMeasurementCampaignCode}

All channel models considered here generate excess attenuation values on top of a site-general channel model. %Throughout our work, 
We use the free-space path loss (FSPL) model as the baseline generic model. The path loss $PL$ in dB at the RX location $s$ is then composed of two parts:
\begin{equation}
	\nonumber PL(s) = FSPL\left[ d(s) \right] + EPL(s) \,\text{,} 
\end{equation}
where $FSPL\left[ d(s) \right]$ is the FSPL in dB at a RX-to-TX distance of $d$ at $s$, and $EPL(s)$ is the excess path loss in dB at $s$. % Note that we have kept the RX location $s$ as inputs for the parameter evaluation procedures to emphasize their site-specific nature. 

\subsection{The AF Propagation Model~\cite{durgin1998measurements}}

The partition-dependent AF propagation model takes advantage of site-specific information by assuming that each instance of one type of obstacle along the LoS path will incur a constant excess path loss. %Only trees were present and hence 
In our case, we counted the number of trees, $N(s)$, along the LoS path to $s$ and added a constant excess path loss in dB, $L_0$, for each of the trees, as follows:
\begin{equation}
	\nonumber EPL(s) = N(s) \cdot L_0 \,\text{.}
\end{equation}

%There are different methods for determining $N(s)$. 
Considering the forest size and the number of RX locations involved, it is extremely difficult and time-consuming to count $N$ at each $s$ on-site. In our work, we simplified the trees, making them vertical lines rather than estimating the cylinder of each tree. Then, the number of trees within the first Fresnel zone for each $s$ was estimated based on manually labeled trunk locations % in the 3D reference system  
and used as the number of obstacle trees. 

\newcommand{\modelPerformFigWidth}{2.7in}
\newcommand{\extraVspaceBetSingleFigAndCaption}{-2mm}
\begin{figure}[t]
	\centering
	\includegraphics[width=\modelPerformFigWidth/560*640]{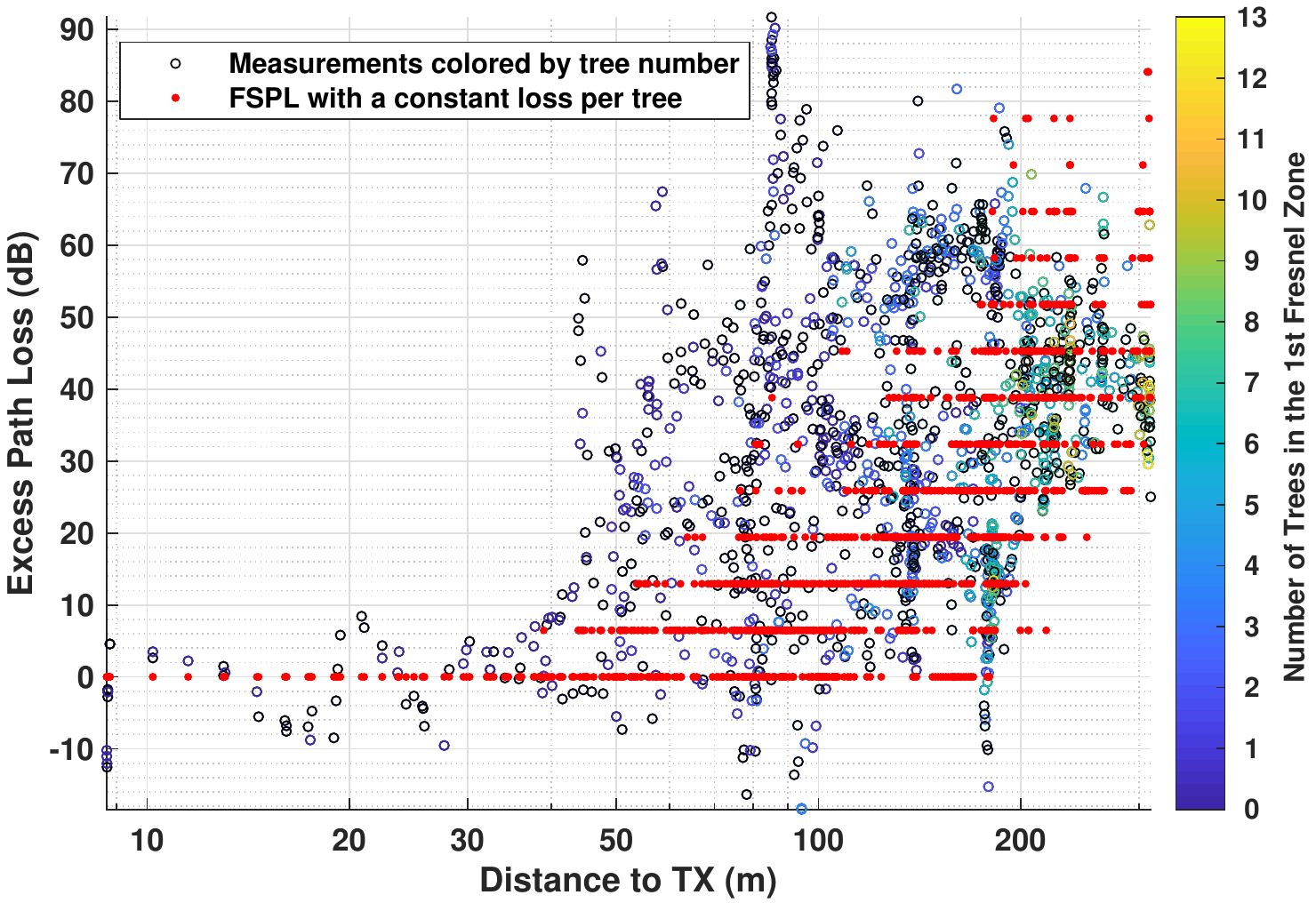} % 4_1_1_MeasAndFreeSpaceExceLossesAndShiftedOverLogDist_Simplified_ColoredByTreeNum.eps => zhang_WCL2018-1396_fig3.eps
    \vspace*{\extraVspaceBetSingleFigAndCaption}
	\caption{The AF propagation model degenerates to a constant-loss-per-tree model in our case. Its predictions fit the shape of the measurement results but have a poor overall accuracy. 
	% As a comparison, a simple shift of the FSPL results overcomes the underestimation issue, but it completely ignores the wide variation of the measurement results.
    }
	\label{model_comparison_partition_dependent}
    \squeezeupafterfigure
\end{figure}

\Cref{model_comparison_partition_dependent} shows the predictions obtained from 
%the constant-loss-per-tree model, to which degenerates in our case
the AF model. The unknown constant $L_0$ was fit according to the measurement data, resulting in a value of 6.47~dB per tree. %To better understand the model's performance, we also plotted another set of predictions from shifting the FSPL values by the mean excess path loss of the whole dataset. 
As can be seen, the AF model closely follows the shape formed by the measurement results. However, it suffers in predicting the correct amount of excess path loss in general. This is expected because 
\hlchanges{
    we have only considered the trunk locations for counting trees, but their physical sizes
    %the leaves and branches
    also play a critical role in attenuating the signal.}
The root mean squared error (RMSE) for the AF model compared with the measurements is 27.96~dB, achieving a 11.47~dB improvement over the FSPL model but still significantly worse than those for the other two empirical models discussed below. %, as suggested by \Cref{tab_overall_perf}. % around 7~dB worse than that for the shifted FSPL predictions, 20.82~dB, even though the latter model ignores the wide path loss variance completely. 
\hlchanges{
    % It is also worth noting that the intuitively simple AF model is very challenging to automate, because counting obstacles of different categories requires intelligence in category definition, obstacle detection, and object classification.
    % Note that it is possible to improve the AF model by differentiating trees of different sizes. However, in our case, trees grow in clusters and it is very challenging to distinguish individual canopies.
    We observe that it may be possible to improve the AF model by classifying trees into different size categories and assigning each category a loss value.  
    However, in our case, trees grew in clusters, making it extremely challenging to distinguish individual canopies and to properly classify trees.}

\subsection{ITU-R Obstruction by Woodland Model~\cite{itu2016vegetation}}

The ITU-R obstruction by woodland model assumes one terminal (the TX or the RX) is located within woodland or similar extensive vegetation, which fits well our measurement scenario. Instead of the number of trees, the ITU model uses the length of the path within the woodland in meters, $d_w(s)$, 
\hlchanges{
which is the distance from the woodland edge to the terminal in the woodland,} % at a given RX location, 
to estimate the excess path loss:
\begin{equation}
\label{equ_itu}
	EPL(s) = A_m \left[1-\text{exp}(-d_w(s)\cdot\gamma / A_m)\right] \text{,}
\end{equation}
where $\gamma\approx 6$~dB/m is the typical specific attenuation for very short vegetative paths at 28~GHz, and $A_m$ is the maximum attenuation in dB. %for one terminal within a specific type and depth of vegetation. 
The most distinguishing feature of this model is the upper limit $A_m$ imposed on the excess path loss.
% caused by the woodland, which agrees with our observations in \Cref{fig_exce_loss_over_tree_num}. 

\begin{figure}[t]
	\centering
	\includegraphics[width=\modelPerformFigWidth]{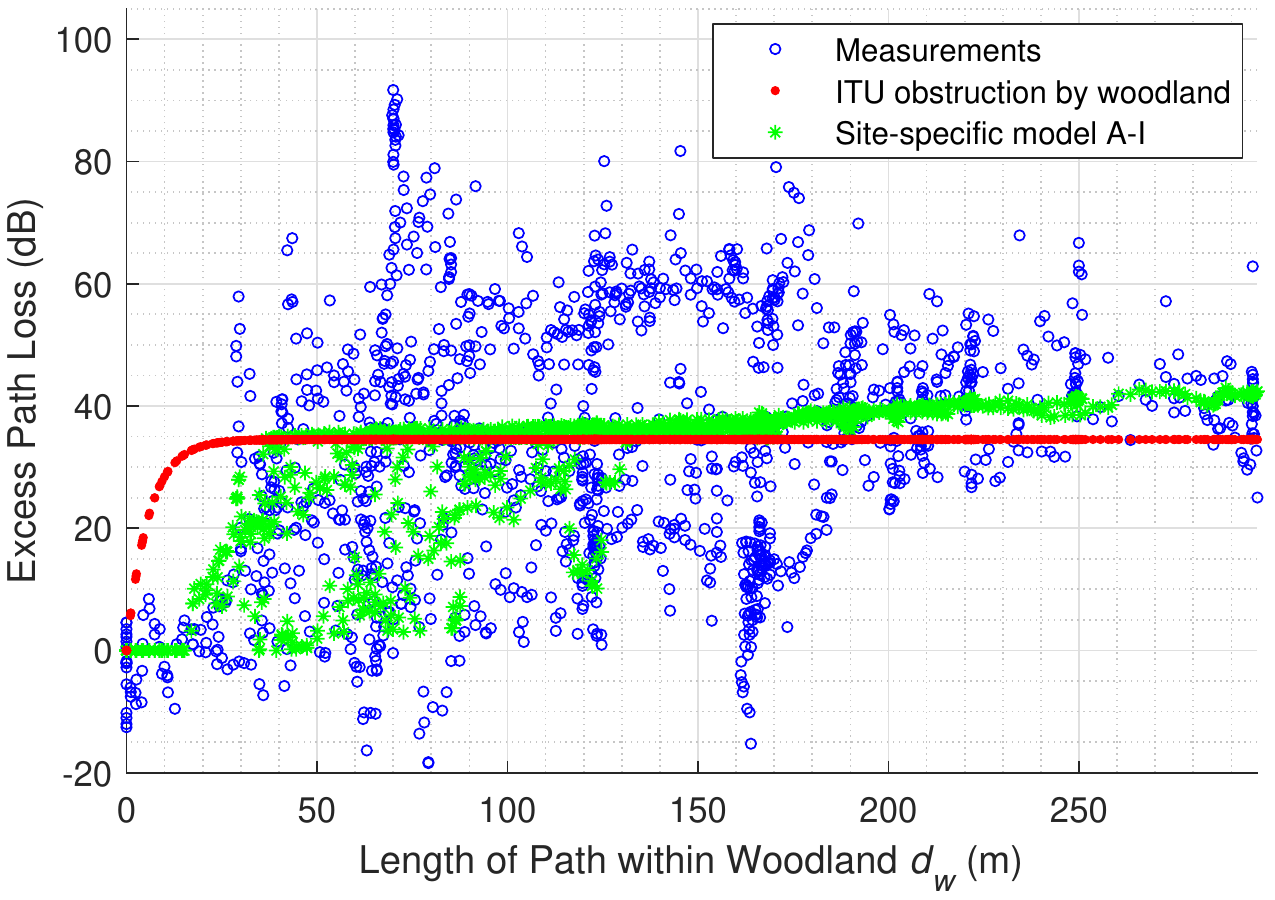} % 7_ItuExcePerformance_MeasVsLengthOfPathInWoodland.eps => zhang_WCL2018-1396_fig4.eps
    \vspace*{\extraVspaceBetSingleFigAndCaption}
	\caption{Predictions from the ITU obstruction by woodland model. As a comparison, the predictions from one site-specific model, which is covered in \Cref{section_site_specific_models}, are also shown. The site-specific model follows the measurements better than the ITU model at the lower and higher ends of $d_w$.}
	\label{fig_itu_results}
    % \squeezeupafterfigure
\end{figure}

Since the TX was installed approximately 15~m away from the forest, this offset has been taken away from the 3D RX-to-TX distance to estimate $d_w(s)$, with the negative values clipped to zero.
%And to avoid negative path length results, $d_w(s)$ is set to be zero for RX locations close to the TX (within the 15~m long range). 
Also, $A_m$ is yet to be determined in \cite{itu2016vegetation} for 28~GHz signals, so we fitted it to our measurement results to obtain the best possible performance, which yielded $A_m \approx 34.5$~dB. The resulting predictions are plotted in \Cref{fig_itu_results}. The ITU model exhibits the best fit among the empirical models considered, with an overall RMSE of 20.08~dB. 
%excels in following the trend of the measurements
However, it clearly overestimates the path loss for locations with $d_w$ smaller than 30~m. At those locations, the LoS path may be clear or blocked by only a couple of trees, differing from a typical woodland blockage scenario. 
On the other hand, the ITU model underestimates the path loss for large $d_w$. 

\subsection{Weissberger's Model~\cite{weissberger1982initial}}

% "applicable to cases in which the ray path is blocked by dense, dry, in-leaf trees"

Weissberger's model, or Weissberger's modified exponential decay (WMED) model, can be formulated as follows:
\newcommand{\horiSpaceToShrink}{-0.8em}
\begin{equation}
	\nonumber EPL(s) = {
		\begin{cases}
			0.45\,f_c^{{0.284}}\,d_f(s)			 &\hspace{\horiSpaceToShrink}{\mbox{, if }}0<d_f(s)\leqslant 14\\
            1.33\,f_c^{{0.284}}\,d_f^{{0.588}}(s) &\hspace{\horiSpaceToShrink}{\mbox{, if }}14<d_f(s)\leqslant 400
		\end{cases}
        }
\end{equation}
where $f_c$ is the carrier frequency, %28~GHz in our case, 
and $d_f(s)$ is the foliage depth in meters along the LoS path for the RX location $s$. The model treats locations with
%deep-enough foliage, that is, 
$d_f(s)$ $>$14 differently from those with less foliage blockage.

% \begin{figure}[t]
% 	\centering
% 	\includegraphics[width=\modelPerformFigWidth]{./eps/8_1_WeissbergerPerformance_MeasVsFoliageDepth.eps}
%     \vspace*{\extraVspaceBetSingleFigAndCaption}
% 	\caption{Predictions from Weissberger's model. Results from \textit{site-specific model A-I} are shown as a reference. Weissberger's model slightly underestimates the loss for RX locations with shallow vegetation blockages and overestimates the loss for those with deep vegetation blockages.}
% 	\label{fig_weiss_results}
% \end{figure}

% \begin{figure}[t]
% 	\centering
% 	\includegraphics[width=\modelPerformFigWidth]{./eps/8_2_SiteSpecModels_MeasVsFoliageDepth.eps}
%     \vspace*{\extraVspaceBetSingleFigAndCaption}
% 	\caption{Predictions from the site-specific models. These models have very competitive performance.}
% 	\label{fig_site_spec_results}
% \end{figure}

We have taken an image processing approach to 
\hlchanges{
    automatically}
obtain the site-specific foliage depth, $d_f(s)$, 
\hlchanges{
    which is the sum total distance for the intersections of the direct path and the foliage regions.}
Both the LiDAR data and the terrain elevation data from the USGS were rasterized onto the same set of reference location points. The foliage regions were then extracted by thresholding their difference, % by 0.5~m, 
resulting in the foliage regions illustrated in \Cref{fig_measurement_overviews}\subref{fig_tree_locaitons_zoomed_in}. Along the LoS path, the ratio of the number of foliage region pixels over the total number of pixels was calculated and multiplied with the corresponding 3D RX-to-TX distance to get the foliage depth for each RX location. 

\Cref{fig_foliage_depth_based_models}\subref{fig_weiss_results} compares the predictions from the WMED model with the measurement results. 
% Foliage depths have been computed via the site-specific geographic information,
% we have to ensure the best possible performance of the WMED model, 
Overall, the WMED model gives a reasonably good RMSE value of 22.19~dB.

\newcommand{\figWidthForTwoInARow}{\modelPerformFigWidth-1mm} %{\modelPerformFigWidth}
\begin{figure}[t]
	\centering
	\subfloat[Results from Weissberger's model\squeezeupbetweensubfigs]{
		\includegraphics[width=\figWidthForTwoInARow]{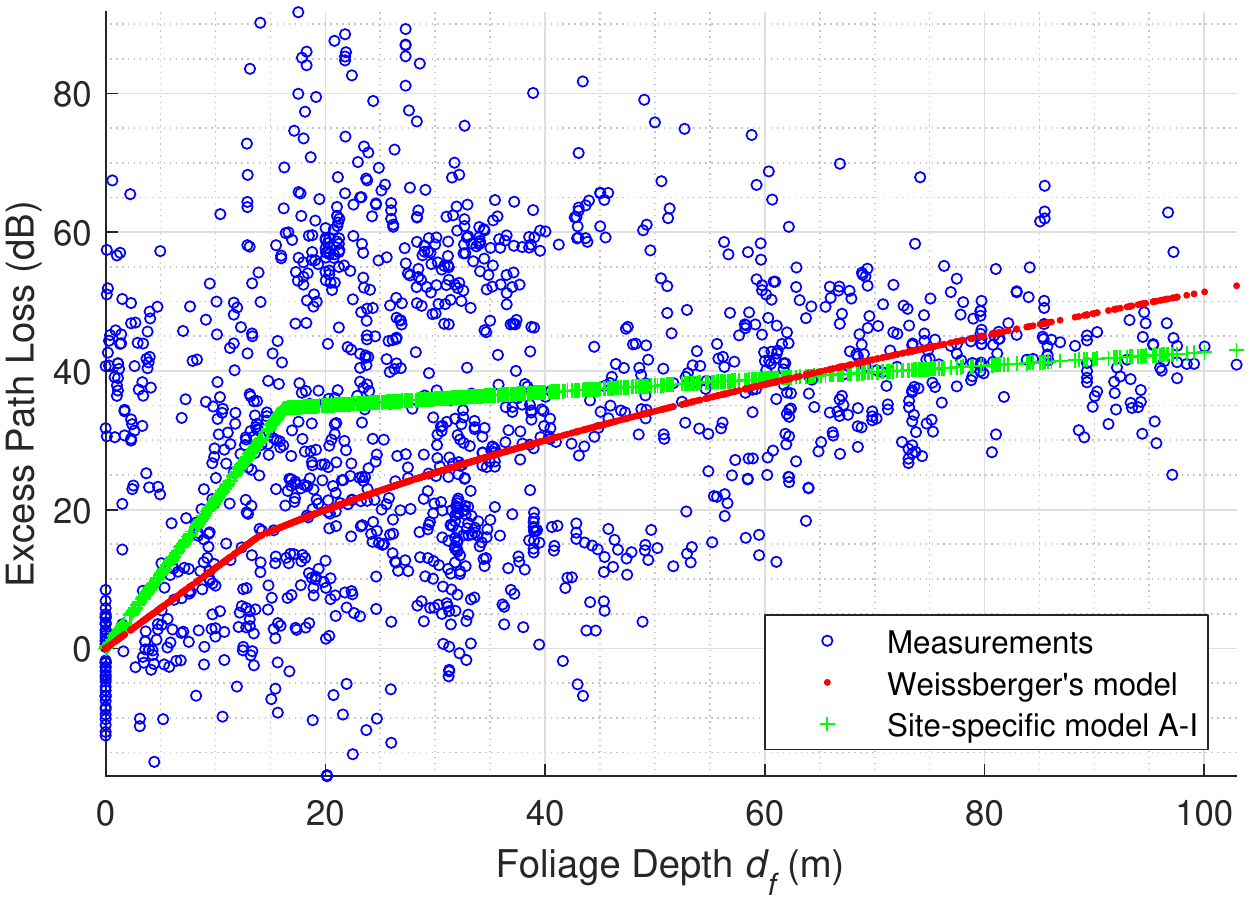} % 8_1_WeissbergerExcePerformance_MeasVsFoliageDepth.eps => zhang_WCL2018-1396_fig5.eps
		\label{fig_weiss_results}
	}
	\hfil
	\subfloat[Results from site-specific models]{
		\includegraphics[width=\figWidthForTwoInARow]{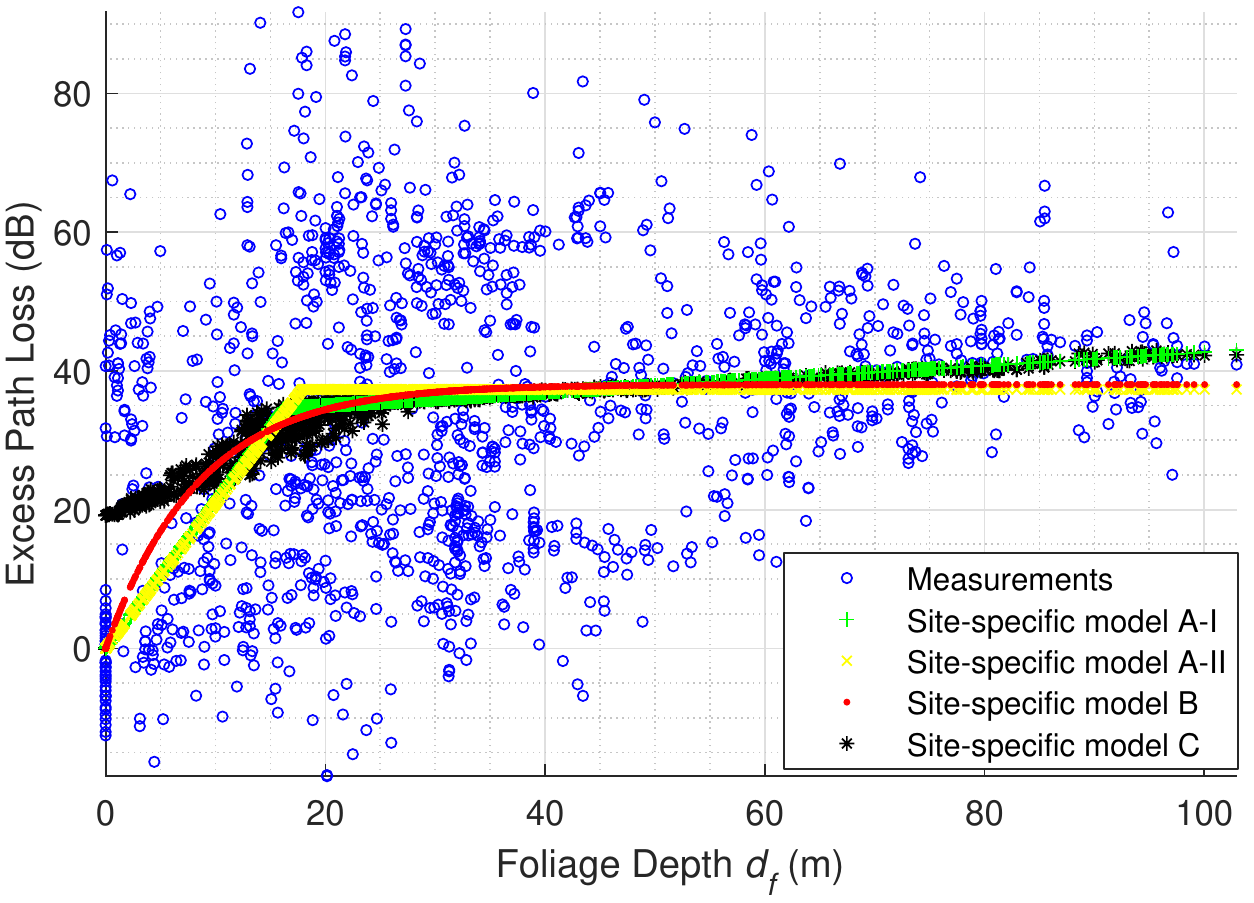} % 8_2_SiteSpecModelsExce_MeasVsFoliageDepth.eps => zhang_WCL2018-1396_fig6.eps
		\label{fig_site_spec_results}
	}
	\caption{\hlchanges{
	Predictions from foliage-depth-based models. Results from \textit{model C}, which makes predictions based on the foliage area in the first Fresnel zone% and allows discontinuity at the origin
	, are also plotted in (b) for comparison.
    % Predictions from foliage-depth-based models. The baseline FSPL values
    % , which have been evaluated individually for all RX locations without considering the foliage, 
    % %for all RX locations
    % are also shown for reference. They form a cloud because the foliage depth, $d_f$, does not increase linearly with the RX-to-TX distance. 
    % (a) Weissberger's model slightly underestimates the loss of RX locations with shallow vegetation blockages and overestimates the loss of those with deep vegetation blockages. To better illustrate this, results from \textit{site-specific model A-I} are shown as a reference. (b) The site-specific models have very competitive performance. Results from \textit{model C}, which makes predictions based on the foliage area in the first Fresnel zone, are also plotted here. 
    }}
	\label{fig_foliage_depth_based_models}
    \squeezeupafterfigure
\end{figure}

\subsection{Site-specific Models}
\label{section_site_specific_models}

Using high-precision publicly available geographic information, existing channel models can be tuned with well-estimated site-specific parameters.
% Using high-precision geographic information that is available publicly,
% %for terrain modeling, 
% such as the LiDAR data, existing channel models can be tuned
% % accordingly 
% with well-estimated site-specific parameters.
% % to perform better using better estimation results for site-specific parameters. 
As a result, simple but powerful site-specific models can be constructed as alternatives. We refer to these as "site-specific" models because their performance depends heavily on 
the accuracy of %  site-specific parameters, which are those 
the parameters evaluated for each site. %, for example, the 3D LoS distance and the foliage depth.
%, given necessary geographic information.

% We have carefully designed these models so that they are fully automatic and thus can be applied in real-life wireless communication networks.

% It is always of high priority to automate the parameter calculation procedure for site-specific models. Otherwise, applying these models will be too tedious to be practically useful. In our work, %besides collecting necessary datasets, 
% some fully-automatic algorithms have been implemented in Matlab for this purpose. Among them, one took 

% \hlchanges{
%     We will use the foliage depth $d_f$, the sum total distance for the intersections of the direct path with the foliage regions, and the foliage area $A_f$, the sum total area for the intersections of the first Fresnel zone with the foliage regions, to construct four site-specific models.}
% This way, the amount of blockage can be accumulated and quantified across the forest in a uniform approach.
By combining the idea of 
\hlchanges{
    evaluating the blockage condition individually for each $s$ from the AF model} 
and the two-slope modeling approach in the WMED model, we constructed \textit{model A-I}:
\renewcommand{\horiSpaceToShrink}{-0.9em}
\begin{equation}
	\nonumber EPL(s) = {
		\begin{cases}
			d_f(s)\cdot L_1 \,\,\text{,}\,\,\,\,\,\,\,\,\,\,\,\,\,\,\,\,\,\,\,\,\,\,\,\,\,\,{\mbox{if }}0\leqslant\,&\hspace{\horiSpaceToShrink}d_f(s)\leqslant D_f\\
            D_f L_1 + \left[ d_f(s)-D_f \right] L_2	\,\,{\mbox{, if }}&\hspace{\horiSpaceToShrink}d_f(s)>D_f
		\end{cases}
        }
\end{equation}
where $d_f(s)$ is the foliage depth in meters at $s$, $L_1$ and $L_2$ are two constants for adjusting the extra loss in dB caused by each meter of foliage, and $D_f$ is the boundary determining when $L_2$ will take effect. The upper bound from the ITU model can be imposed by setting $L_2=0$ to form \textit{model A-II}:
\renewcommand{\horiSpaceToShrink}{-0.8em}
\begin{equation}
	\nonumber EPL(s) = {
		\begin{cases}
			d_f(s)\cdot L_1 	&\hspace{\horiSpaceToShrink}{\mbox{, if }}0\leqslant d_f(s)\leqslant D_f\\
            D_f \cdot L_1 	&\hspace{\horiSpaceToShrink}{\mbox{, if }}d_f(s)>D_f
		\end{cases}
        }
\end{equation}
%where a maximum excess loss will come into force for deep enough foliage blockage, determined by the threshold $D_f$. 

We also reused the ITU model in \Cref{equ_itu} with site-specific foliage depth to form \textit{model B}. That is, $d_f(s)$ is used instead of $d_w(s)$, and parameters $A_m$ and $\gamma$ are set according to the measurements. 

For a fair performance comparison for these three models, we used the WMED boundary $D_f = 14$ for \textit{model A-I} to leave only two adjustable parameters. After fitting these models to our data, we found $L_1 \approx 2.39$~dB/m and $L_2 \approx 0.12$~dB/m for \textit{model A-I}, $L_1 \approx 2.09$~dB/m and $D_f \approx 17.87$~m for \textit{model A-II}, along with $A_m \approx 38.04$~dB and $\gamma \approx 4.47$~dB/m for \textit{model B}. 
%, both are close to the results from the original ITU model. 
The resulting predictions are plotted in \Cref{fig_foliage_depth_based_models}\subref{fig_site_spec_results}. The corresponding RMSE
%and mean absolute error (MAE) 
values are summarized in \Cref{tab_overall_perf}, together with those for the traditional models as references. Note that the site-specific models perform very similarly, and each unit of foliage depth tends to contribute less to the excess loss as foliage depth grows. \textit{Model A-I} does not limit the excess loss as the other two site-specific models do, but it performs slightly better than \textit{model A-II} in terms of RMSE. Overall, \textit{model B} performs the best, but computationally, it is more demanding because of its exponential form. 

We can further push the best RMSE performance to 19.18~dB with \textit{Model C}:% by allowing discontinuity at the origin:
\begin{equation}
	\centering
	\nonumber EPL(s) = {
		\begin{cases}
            0 \,\,\text{,}\,\,\,\,\,\,\,\,\,\,\,\,\,\,\,\,\,\,\,\,\,\,\,\,\,\,\,\,\,\,\,\,\,\,\,\,\,\,\,\,\,\,\,\,\,\,\,\,\,\,\,\,\,\,\,\,&{\mbox{if }a_f(s)=0} \\
			a_f(s)\cdot L_1 + L_0 \,\,\text{,}\,\,\,\,\,\,\,\,\,\,\,\,\,\,\,\,\,\,\,\,\,\,\,\,&{\mbox{if }}0<a_f(s)\leqslant A_f\\
            A_f L_1 + \left[ a_f(s)-A_f \right] L_2	\,\,\text{,}&{\mbox{if }}a_f(s)>A_f
		\end{cases}
        }
\end{equation}
\hlchanges{where foliage area $a_f(s)$ is the sum total area for the intersections between the first Fresnel zone at RX location $s$ and the foliage regions;} 
$L_0$ (dB), $L_1$ (dB/m$^2$), and $L_2$ (dB/m$^2$) are constants adjusting the excess loss contribution; and $A_f$ is the boundary determining when the foliage is deep enough for $L_2$ to take effect. According to our measurement results, we have $L_0 \approx 19.14$~dB, $L_1 \approx 2.09$~dB/m$^2$, $L_2 \approx 0.06$~dB/m$^2$, and $A_f \approx 18.02$~m$^2$. This model has a sudden jump at the origin. Its prediction results are also shown in \Cref{fig_foliage_depth_based_models}\subref{fig_site_spec_results} for reference.

\begin{figure}[t]
	\centering
	\subfloat[Regional RMSE improvement over the ITU model\squeezeupbetweensubfigs]{
		\includegraphics[width=\figWidthForTwoInARow]{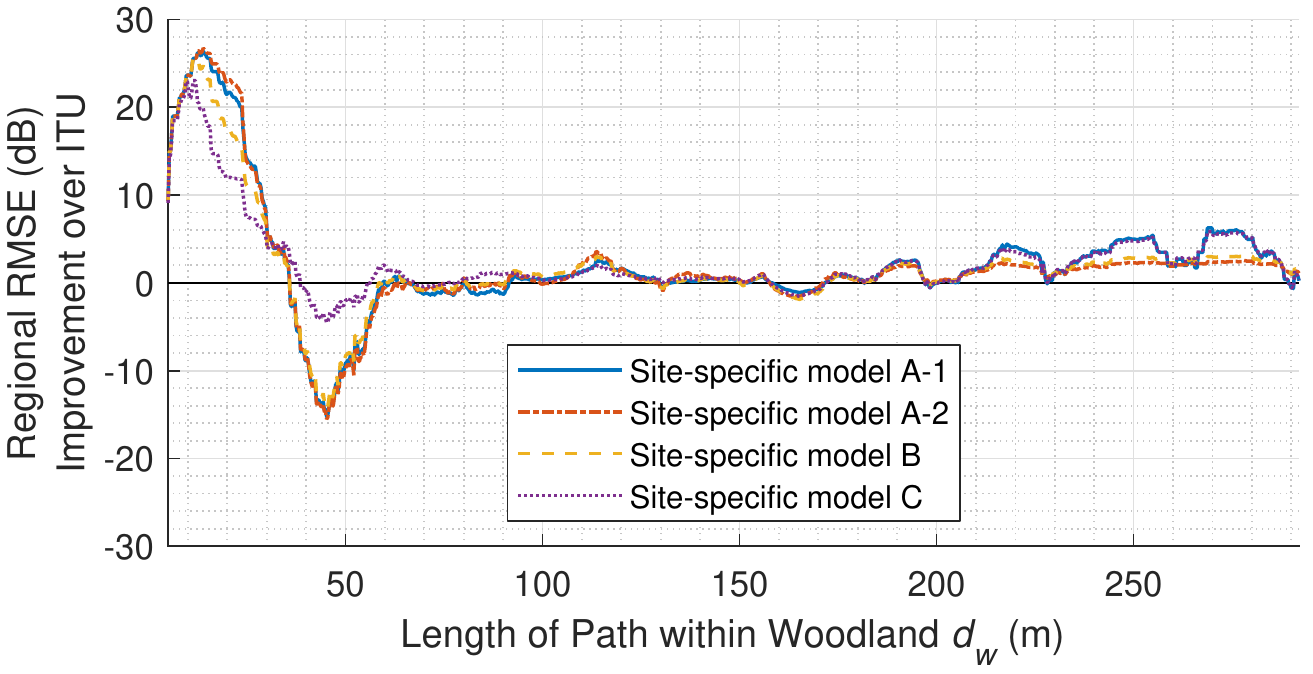} % 9_1_RegionalRmseImprovementOverItu.eps => zhang_WCL2018-1396_fig7.eps
		\label{fig_regional_perf_itu}        
	}
	\hfil
	\subfloat[Regional RMSE improvement over the WMED model]{
		\includegraphics[width=\figWidthForTwoInARow]{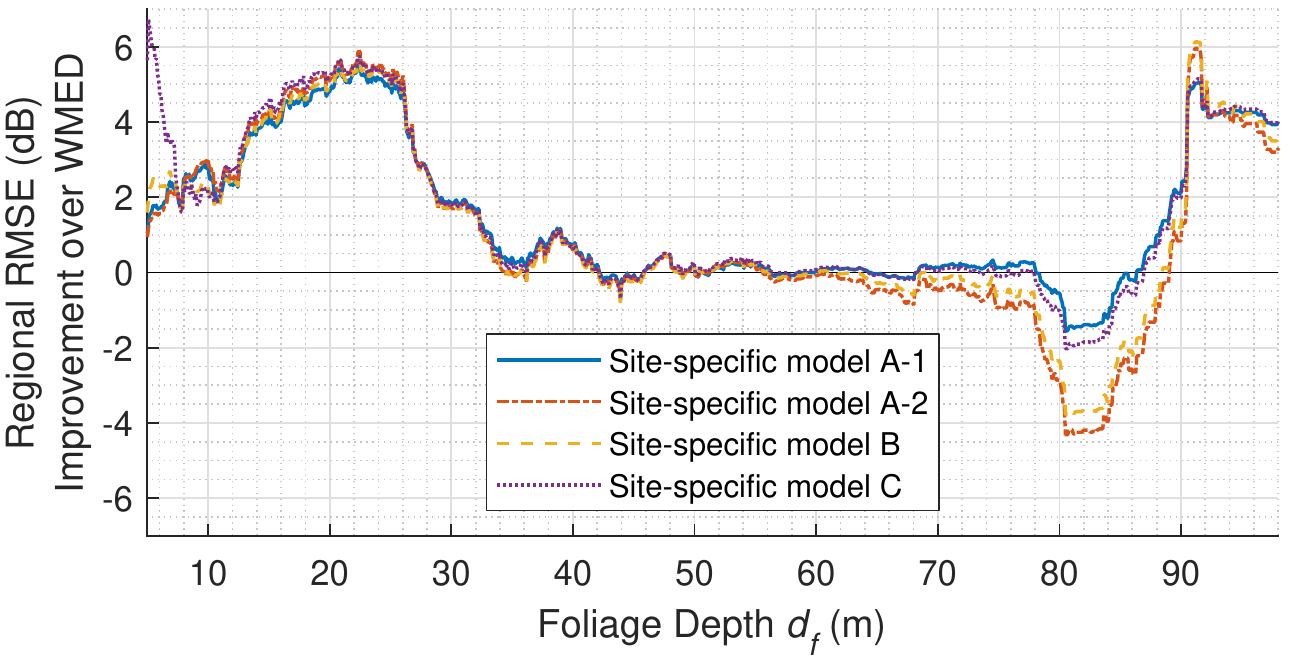} % 9_2_RegionalRmseImprovementOverWeis.eps => zhang_WCL2018-1396_fig8.eps
		\label{fig_regional_perf_wmed}
	}
	\caption{Regional performance improvement for site-specific models using a window size of 10~m. (a) Compared with the ITU model, site-specific models work significantly better for locations close to the TX and reasonably better for those far away. However, models \textit{A-I}, \textit{A-II}, and \textit{B} suffer a severe performance degradation for $d_w$$\in[35,  60]$~m, which is less of an issue for \textit{model C}. (b) Compared with the WMED model, site-specific models again work reasonably better for extreme cases. A performance deterioration is observed at a foliage depth of around 80~m, where models \textit{A-I} and \textit{C} are less influenced.}
	\label{fig_regional_perfs}
    \squeezeupafterfigure
\end{figure}

\hlchanges{
    The most important feature for these models is that 
    they are fully automatic and thus can be applied in large-scale wireless communication networks. 
    Site-specific information was fetched from Google and USGS servers. Foliage information was extracted, and channel modeling performed, by our automated algorithms.}
    %Site-specific information was automatically fetched, on-demand if necessary, from Google and USGS, and algorithms have been implemented to extract foliage regions and carry out channel modeling tasks for different RX sites.}
Another advantage of our site-specific models is their consistently good performance throughout the whole dataset. 
To demonstrate this, regional RMSE improvements over the ITU and WMED models are evaluated in terms of $d_w$ and $d_f$, respectively, as summarized in \Cref{fig_regional_perfs}. For our dataset, the ITU model works very well, as shown in \Cref{tab_overall_perf}. However, according to \Cref{fig_regional_perfs}\subref{fig_regional_perf_itu}, the ITU model suffers from an RMSE degradation of as much as 20~dB compared with the site-specific models in the low-vegetation-coverage region ($d_w$$<$30~m).  
For large $d_w$, this value is observed to be as much as 6~dB. A visual comparison for predictions from the ITU model and \textit{model A-I} is provided in \Cref{fig_itu_results}, where \textit{model A-I} clearly works better for extreme cases at the low and high ends of $d_w$. Similar comparisons have been carried out for the WMED model in \Cref{fig_foliage_depth_based_models}\subref{fig_weiss_results}  and \Cref{fig_regional_perfs}\subref{fig_regional_perf_wmed}. The WMED model slightly underestimates the path loss at RX locations with a small $d_f$ and overestimates it at 
%those with a 
large $d_f$. 
% It is worth noting that empirical models do work significantly better in some distance regions, whereas \textit{model C} suffers less in those regions. 
% Another performance metric, the mean absolute error (MAE), has also been considered in \Cref{tab_overall_perf}, which reveals similar results as the RMSEs.

\newcommand{\tableArrayStretch}{1.1}
\begin{table}[t]
	\renewcommand{\tabcolsep}{2.2pt}%{8.5pt}
	\renewcommand{\arraystretch}{\tableArrayStretch}
    \squeezeupbeforetable
	\caption{Overall Performance}
	\label{tab_overall_perf}
	\centering
    \begin{tabular}{|>{\bfseries}c||c|c|c|c|c|c|c|c|}
    \hline
                       & \textbf{Baseline} & \multicolumn{3}{c|}{\textbf{Traditional}} & \multicolumn{4}{c|}{\textbf{Site-Specific}} \\ \hline
    \textbf{Model}     & FSPL 	  & AF  & ITU   & WMED & \textit{A-I}  & \textit{A-II}  & \textit{B}  & \textit{C} \\ \hline
    \textbf{RMSE (dB)} & 39.43 	  & 27.96 & 20.08 & 22.19       & 19.96 & 20.02 & 19.93 & 19.18 \\ \hline
    %\textbf{MAE (dB)}  & 34.09 	  & 21.95 & 16.42 & 16.82       & 15.89 & 16.07 & 16.08 & 15.51 \\ \hline
    \end{tabular}
    \squeezeupaftertable
\end{table}

\section{Conclusion}

\hlchanges{
    A comprehensive channel model comparison for attenuation through vegetation was conducted using measurements in a coniferous forest near Boulder, Colorado. Inspired by the results, we developed novel site-specific models for consistent improvement in prediction accuracy through shallow to deep vegetation blockages. They are fully automatic, easy to implement, and feasibly applicable to machine learning frameworks. 
% A comprehensive channel model comparison for attenuation through vegetation was conducted using data from measurements in a coniferous forest near Boulder, Colorado. The partition-dependent AF model is intuitive and site-specific but hard to automate with satisfying performance. The ITU obstruction by woodland model works extremely well overall, but only for scenarios with a moderate amount of foliage blockage. Weissberger's model is similar. By merging the existing models, site-specific models are proposed for consistent performance through shallow to deep vegetation blockages. These models are easy to implement and adaptable to machine-learning frameworks. 
}

%\balance

% ============== Refs ==============
\nocite{}
\bibliographystyle{IEEEtran}
\bibliography{zhang_WCL2018-1396} % ieeeWirelessLetter2018 => zhang_WCL2018-1396

\end{document}